\documentclass[letterpaper, twocolumn, 10pt]{article}
\usepackage{graphics}
\usepackage{usenix}

\usepackage{url}
\usepackage{verbatim}
\usepackage{subfigure}
\usepackage{amssymb}
\usepackage{graphicx}
\newcommand{\msg}[1]{{\small \textsf{#1}}}

\newcommand{\eat}[1]{}

\begin{document}

\date{}
\title{\Large \bf Transparent Format Migration of Preserved Web Content }

\author{
David S. H. Rosenthal
\and
Thomas Lipkis
\and
Thomas Robertson
\and
Seth Morabito\\
\small{Stanford University Libraries, {CA}}
}

\maketitle

%For the first page - it's separate from the rest in terms of page style.
\thispagestyle{empty}

\begin{abstract}

The LOCKSS digital preservation system collects content by crawling the
web and preserves it in the format supplied by the publisher.
Eventually,
browsers will no longer understand that format.
A process called \emph{format migration} converts it to a newer
format that the browsers do understand.
The LOCKSS program has designed and tested an initial implementation
of format migration for Web content that is transparent to readers,
building on the content negotiation capabilities of HTTP.

%{\bf\verb!$Revision: 1.15 $!}
\end{abstract}

\section{Introduction}
\label{sec:introduction}

Eventually, any format in which digital content might be stored will
become obsolete.  A format is said to be obsolete when current hardware and
software are no longer able to render information represented in it
understandable to readers. The design of digital preservation systems must
anticipate this obsolesence, and incorporate a strategy by which the
content they preserve will still be understood by readers after
multiple generations of formats have become obsolete.
Two such strategies have been identified:
\begin{description}
\item [Emulation] in which the content is both preserved and presented
to readers in the original format~\cite{Rothenberg1995}.
\item [Migration] in which the content is presented in a current format;
it may be preserved in a succession of current formats or in the original
format which is transformed on request into the current format for
presentation~\cite{Ockerbloom1998}.
\end{description}

Some software business models depend on a rapid upgrade
cycle.  In these areas rapid format change is normal;  users who do
upgrade produce a format users who have yet to upgrade cannot interpret.
This is a powerful motivation for further upgrades, and thus a powerful
income generator.  But note that rapid format
\emph{change} doesn't imply rapid format \emph{obsolescence}.  An
upgraded application that didn't accept old formats would not be
an effective income generator.

We provide an overview of the problem of format obsolescence as applied
to Web content and,  in this context, examine possible implementations
of the two strategies.  We identify the practical difficulties that face
any implementation of emulation;
they led us to choose the migration strategy.
We describe the design and implementation of a transparent,
on-access format migration capability for the
LOCKSS\footnote{LOCKSS is a trademark of Stanford University.
It stands for Lots Of Copies Keep Stuff Safe.} system for preserving Web
content.

Our implementation is capable of transparently presenting content collected
in one Web format to readers in another Web format,
with no changes needed to browsers.
The reader need take no special action to cause this to happen,
nor even be aware that it is happening.
This appears to be the first time that a production digital
preservation system has demonstrated transparent format migration
of live content collected from the Web for end-users.

\section{Format Obsolescence of Web Content}
\label{sec:FormatObsolescence}

A Web format may be said to be obsolescent when widely-used browsers
are no longer able to present content in the format to their
readers.

To the casual observer it may appear that the format in which Web
content is supplied is solely determined by the Web server,  possibly
by the file name extension.  In fact,  the format is determined by
one of a set of mechanisms for \emph{content negotation} defined
in Section 12 of~\cite{RFC2616} which are capable of negotiating format,
language, and encoding.  The mechanism for format negotation uses the
\msg{Accept:} header defined in Section 14.1 of~\cite{RFC2616}.
A browser sends this to the server with a list
of acceptable \msg{Mime-Type} values,  each with a numeric preference
value between 0 and 1.  If it is capable of supplying the requested
content in multiple formats the server \msg{MAY} decide on the basis
of this list and its preference values which format to supply.
Browsers determine the format in which the server has decided to
supply the content they requested using the \msg{Mime-Type:} header.

In practice,  a browser does not know when it issues a request
for a URL whether it refers to text,
audio, video or some other class of object.  Browsers therefore
send a default \msg{Accept:} header on most \msg{GET} requests
specifying their preferred \msg{Mime-Type} values.  These lists
typically include a low-preference default \msg{*/*;q=0.1} saying,
in effect, ``if you can't give me what I want, give me what you have''.
Because browsers indicate in this way their willingness to receive
\emph{any} format,  there is some difficulty in determining when
obsolescence has ocurred.  

In brief,
the problem of format obsolescence for a system preserving Web content is
that of what to do when it receives a request for some content that was
collected in format \msg{F/G}, say \msg{image/gif},
whose \msg{Accept:} header indicates \msg{F/G} is not an acceptable format.

In the light of the \msg{*/*;q=0.1} usage, there are two ways in
which this can happen;
the browser can explictly signal it or the server can be configured
to assume it.
Although Section 12 of~\cite{RFC2616} does not define the semantics of
a 0 preference value,  it appears that servers treat this as
an instruction not to send content in that \msg{Mime-Type}.
Thus even a browser that uses \msg{*/*;q=0.1} can flag a format
as unacceptable by \msg{F/G;q=0}.
Alternatively,
a server could be configured not to recognize \msg{F/G} as matching
\msg{*/*}.

Fortunately,  since Web browsers and their plug-ins are normally free,
there are few incentives for rapid format change and particularly
obsolescence.  No-one clamors to remove support for an old Web format;
it is valuable so long as there is content on Web servers that has
not yet been migrated to a current format.  There are no good
ways to motivate small Web sites to perform this migration,
so old Web formats die a very slow death.  From the viewpoint of
digital preservation,  this makes Web content easier to handle;
there will be plenty of time to implement a format migration.

\subsection{Emulation of Obsolete Web Formats}
\label{sec:FormatObsolescence:Emulation}

The goal of the \emph{emulation} preservation strategy is to avoid
the loss of fidelity that is likely to result from converting content
from one format to another.
If the content is preserved in its original format and presented to the reader
in that format, no conversion is needed.
What is needed is the ability of a future reader to run the software the
original reader would have run to experience the content.
The emulation strategy seeks to provide that by preserving the original
software as well as the content,  and providing the future reader with
a software emulation of the environment needed to run the original
software to interpret the preserved content in its original format.
In a suitable context the emulation strategy is attractive;
it is being pursued,
for example, by a collaboration between IBM~\cite{Lorie2004} and
the Koninklijke Bibliotheek (KB, Dutch National Library)~\cite{Wijngaarden2004}
which has built a PDF interpreter that runs on a Universal Virtual
Computer (UVC, a virtual machine designed to be easy to port to future
environments).  The terms under which the KB preserves content 
make this appropriate; they mandate that it be accessed only at the KB,
where deployment of the UVC is easy.

In the Web context emulation means that a future reader wishing
to read a preserved Web page that contains some content in an
obsolete format must somehow find out the approximate date of
the original content,  then locate a preserved browser or plug-in
of that date,  and the appropriate emulation needed to allow that preserved
browser or plug-in to run in the reader's current computing environment.
The reader must then invoke this emulation to run the preserved
browser or plug-in to view the Web page.

Since in the emulation strategy all these activities take place
in the reader's environment,  there is little the preservation
system can do to enable them.  It has no control over the reader's
environment.  Indeed,  if it is disseminating the preserved content
by acting as a Web server,  it will have almost no knowledge of the
reader's enviroment.  Although the effect of a successful emulation
strategy would be to prevent the preservation system ever seeing
a request with \msg{F/G;q=0} in its \msg{Accept:} header,
the practical difficulties in implementing both the emulation of
instruction sets, operating systems, etc. and in deploying both the
approrpiate emulation and the appropriate preserved browser or plug-in to
the appropriate reader are formidable.

\subsection{Migration of Obsolete Web Formats}
\label{sec:FormatObsolescence:Migration}

\emph{Migration} of Web content from an obsolete format to a current one
can take place at any time between the point at which the content is
collected to the point at which the reader requests access to it.
We examine three points that have been implemented,
from the earliest to the latest.

\subsubsection{Migration On Ingest}
\label{sec:FormatObsolescence:Migration:Ingest}

The National Archives of Australia (NAA), faced with a requirement to preserve
vast volumes of government information in a wide variety of mostly
proprietary formats, chose a strategy of
\emph{migration on ingest}~\cite{Heslop2002}.
They pre-emptively migrate the content they receive into one of a small
number of carefully chosen formats before preserving it. If their choices
turn out well,  this pragmatic approach has significant advantages:
\begin{itemize}
\item It can postpone the need for future migration for a long time,
allowing both economies in operation and the use of better,
future technology for performing the next migration.
\item It can greatly reduce the cost of eventual future migrations
by reducing the number of formats to be migrated.
\end{itemize}

Both these advantages are greatly enhanced if the formats chosen are
open standards and are supported by open source software, as they are
in the case of NAA.  Most of the material NAA handles is not from
the Web,  and most Web formats would meet their criteria without
an initial migration.

The disadvantages of this approach are two-fold. First, it does not
fully satisfy the requirements of archivists, because the content is
not preserved in its original form~\footnote{NAA actually does preserve
both the original and the migrated format but expects that
access to the original will be an exception}.
Some potentially useful information
may be lost in the initial migration.  Second, it postpones the format
migration problem but does not actually solve it.  Even the chosen
formats will eventually become obsolete.

\subsubsection{Batch Migration}
\label{sec:FormatObsolescence:Migration:Batch}

When a format in which some content is being preserved is thought likely
to become obsolete,  a \emph{batch migration} process can be preemptively
undertaken.
The preserved content in the obsolete format is converted to a current
format \emph{en masse}.  Some stand-alone tools for doing
so have been developed~\cite{Walker2004} but they have
yet to be integrated into a complete digital preservation system.  When
a reader requests access to the preserved content,  the result of the
conversion will be delivered.

The DAITSS (Dark Archive In The Sunshine State) is designed to use
a batch migration strategy~\cite{DAITSS2004}~\footnote{DAITSS also preserves
both the original and the migrated format}.
As a dark archive,  one not intended to
be accessed by readers but maintained in a controlled environment,
this is an appropriate solution.
The archive has total control of the environment and no urgent
demands from end-users to satisfy.

\subsubsection{On-Access Migration}
\label{sec:FormatObsolescence:Migration:OnAccess}

The alternative migration approach,
\emph{migration on access},
postpones format migration until the reader actually requests the
preserved content.  It avoids the disadvantages of the
other migration strategies by preserving the content in its
original formats.
When a format is thought likely to become obsolete,
the digital preservation system is enhanced with the ability to
present the reader, upon request, with the requested content in
a current format.
In effect,  the migration tool is integrated into the dissemination
pipeline of the preservation system rather than being applied to the preserved
content.

This approach requires the ability to convert dynamically from the obsolete
to the current format,
but it offers significant advantages:
\begin{itemize}
\item Content is preserved in its original format, satisfying the archivists'
requirements and avoiding the risk of information loss from buggy
format convertors.  This risk is clearly enough to motivate systems
using other migration strategies to hedge their bets by preserving
the original format too.
\item Preserved content is migrated by the most recent, and presumably best,
technology available at the time the reader requests access.
\item Preserved content is rarely accessed, thus delaying format migration
until it is actually required reduces the resource cost of the process
by the proportion that is never accessed,
and by the decreasing cost of technology through time.
\item Content can be migrated directly from the original to the current format,
minimizing the effects of format conversion artefacts.
\item The format converters, once developed, can themselves be preserved
to document the original format.  Note that a converter can be developed
pre-emptively \emph{before} the format goes obsolete and preserved against
future need if the format's longevity is suspect.
\item As with other migration strategies, careful choice of the format to
migrate to can greatly reduce the need for and cost of future migrations.
\end{itemize}

The disadvantages of the migration on access strategy are that dynamic
format migration may impose significant delays on reader's accesses to
preserved material, and that it requires close integration with the
dissemination pipeline delivering the digital preservation system's
preserved content to its readers.

\section{Format Migration in the LOCKSS System}
\label{sec:LOCKSS}

The LOCKSS system provides librarians with a simple, low-cost tool they
can use to ensure their community's continuing access to material
published on the Web~\cite{Maniatis2003lockssSOSP}.
It is designed to handle both for-fee subscription e-journals
and open access material whose copyright is held by the publisher,
not the library's institution.  Libraries run LOCKSS peers, low-cost PCs
running free, open source software that:
\begin{description}
\item [Collects] the material to be preserved by crawling the publisher's
web site, after verifying that the publisher has granted suitable permission.
\item [Preserves] the material by cooperating with other peers holding
the same material in a mutual audit process by running \emph{polls}
to identify any missing or damaged content and repair it.
\item [Disseminates] the preserved material by acting as a proxy cache,
intercepting requests from the library's browsers for the original URL
from which the material was collected.  If the publisher's copy is
still available,  it is delivered.  Otherwise the preserved copy is
delivered.
\end{description}

The LOCKSS system was released for production use in April 2004 and about
80 libraries world-wide now use it. Publishers of over 2000 titles have
endorsed the system.

\subsection{Design}
\label{sec:LOCKSS:Design}

In the LOCKSS system it is natural to use the migration on access strategy.
Preserving content in the original format greatly simplifies the mutual
audit and repair process,  and the LOCKSS system already implements the
complete dissemination pipeline into which the migration process must be
integrated.

The LOCKSS system is being enhanced to provide:
\begin{itemize}
\item An API for plug-in format convertors,  by which they can register
their input and output \msg{Mime-Type} values,  and by which the
LOCKSS web proxy code can invoke them to perform on-the-fly conversion.
\item A matching process which takes the \msg{Accept:} header of
incoming requests and compares it to the original format of the
preserved content.  If the original format is not acceptable,
the matching process searches the table of registered format convertors
looking for one which takes the original format as input and whose
output format is acceptable.  If a suitable convertor is not found,
a 406 error is returned as required by Section 10.4.7 of~\cite{RFC2616}.
\item A distributed registry of convertors,  similar to the distributed
registry of the plug-ins that adapt the LOCKSS system to particular
content.  These registries treat Java classes exactly as other Web
content; collecting them by crawling the Web and preserving them by mutual
audit and repair.
\end{itemize}

\subsection{Proof-of-Concept Implementation}
\label{sec:LOCKSS:Implementation}

To confirm the feasibility of this design,  a proof-of-concept was implemented
and tested.  
We chose an ``obsolete'' format widely used in actual content collected
by the production LOCKSS system,  and a suitable ``current'' format
to replace it.  The obsolete format was GIF~\cite{GIFspec},
an old format for images.
% It forms about XXX\% of the files collected
% by a typical LOCKSS system in production.
The GIF format has been
deprecated by many open source advocates for reasons connected with
intellectual property restrictions,  and they have developed the
PNG~\cite{PNGspec} format as a replacement.  This background makes
our assessment less artificial,  as the format migration in question
has been actively solicited.  Tools for converting
from GIF to PNG are widely available,  as would be expected if a
widely used Web format were to become obsolete.

We did not implement the full \msg{Mime-Type} matching
process,  but rather a configuration option that prevented
\msg{image/gif} from matching any \msg{Accept:} header.
The mis-match triggered a GIF-to-PNG conversion directly,
delivering the content converted to PNG at the original URL
but with \msg{Mime-Type=image/png}.

\subsection{Assessment}
\label{sec:LOCKSS:Assessment}

\begin{figure}
\centering
\includegraphics[width=8cm]{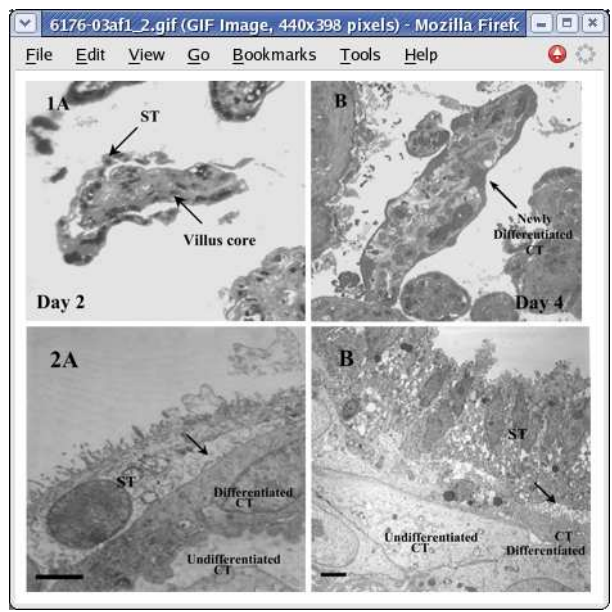}
\caption{A GIF image from an article in \emph{Journal of Histchemistry and Cytochemistry}
before the simulated obsolescence of GIF. Note the header.}
\label{fig:before}
\end{figure}

\begin{figure}
\centering
\includegraphics[width=8cm]{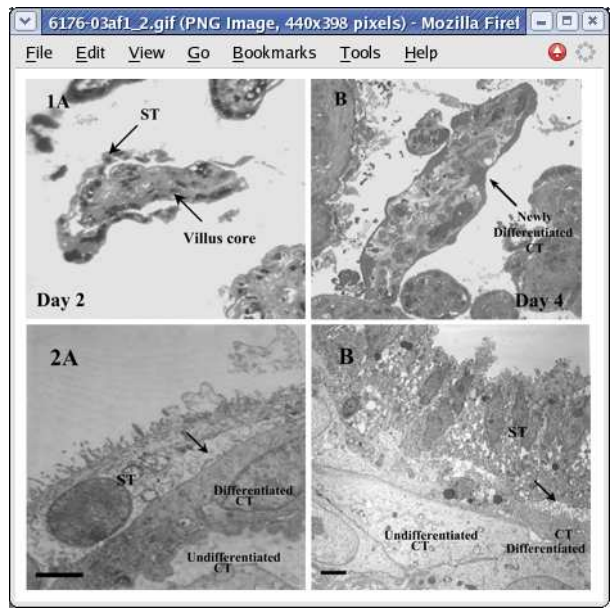}
\caption{The same image after the simulated obsolescence of GIF.
The header shows that the GIF file has been converted to PNG.}
\label{fig:after}
\end{figure}

As can be seen from the before (Figure~\ref{fig:before}) and after
(Figure~\ref{fig:after}) screen-shots,  this format migration is
not perceptible to the user. Nor does GIF-to-PNG format migration
incur a noticeable delay in accessing the page.

\section{Future Work}
\label{sec:FutureWork}

The next step is to replace the proof-of-concept implementation
by a full implementation of the API for plug-in format convertors,
and an broader set of convertors than just GIF-to-PNG.  This
implementation will need a more realistic-scale test,  and we
are arranging to conduct one.

Another approach would be to connect the API to a format migration
service such as TOM (Typed Object Model)~\cite{Ockerbloom1998}.
We are investigating this possibility.

The development of future format convertors will be significantly easier
if more,  and more reliable format metadata is available.  We are working
towards incorporating Harvard's JHOVE~\cite{JHOVE} format metadata
extraction and validation technology to improve the quality of the format
metadata in the LOCKSS system.

\section{Conclusion}
\label{sec:Conclusion}

We have designed, implemented a proof-of-concept and demonstrated transparent
format migration on access
for the LOCKSS digital preservation system.  By doing so we have validated
one of the possible format migration strategies,  and reassured the
community of LOCKSS users that when the time comes the content they are
preserving will remain accessible despite the obsolescence of the formats
in which it was collected.

\section{Acknowledgments}
\label{sec:acknowledgments}
We are grateful to Claire Griffin,
and to Vicky Reich, the director of the LOCKSS
program, for their help.

This work is supported by the National Science Foundation (Grant No.\
9907296) and by the Andrew W. Mellon Foundation.
Any opinions, findings, and conclusions or
recommendations expressed here are those of the authors and do not
necessarily reflect the views of these funding agencies.

{\footnotesize \bibliographystyle{acm} \bibliography{../common/bibliography}}

\end{document}